\documentclass[pdflatex,sn-basic]{sn-jnl}

\jyear{2021}%

\theoremstyle{thmstyleone}%
\theoremstyle{thmstyletwo}%

\theoremstyle{thmstylethree}%

\def\pa{\partial}
\def\vb{{\bf v}}

\def\Bb{{\bf B}}
\def\vf{{\bf v}_m}
\def\Bf{{\bf B}}

\def\ep{{\bf e}_{\phi}}

\def\MC{meridional circulation}
\def\DF{differential rotation}

\begin{document}

\title[Flux transport dynamo model]{The emergence and growth of the flux transport dynamo model
of the sunspot cycle}


\author*[1]{\fnm{Arnab} \sur{Rai Choudhuri}}\email{arnab@iisc.ac.in}

\affil*[1]{\orgdiv{Department of Physics}, \orgname{Indian Institute of Science}, 
\city{Bengaluru}, \postcode{560012}, \state{Karnataka}, \country{India}}


\abstract{The sunspot cycle is the magnetic cycle of the Sun produced by the dynamo process.  A central
idea of the solar dynamo is that the toroidal and the poloidal magnetic fields of the Sun
sustain each other.  We discuss the relevant observational data both for sunspots (which
are manifestations of the toroidal field) and for the poloidal field of the Sun. We point
out how the differential rotation of the Sun stretches out the poloidal field to produce the
toroidal field primarily at the bottom of the convection zone, from where parts of this toroidal
field may rise due to magnetic buoyancy to produce sunspots. In the flux transport dynamo 
model, the decay of tilted bipolar sunspot pairs gives rise to the poloidal field by the
Babcock--Leighton mechanism. In this type of model, the meridional circulation of the Sun,
which is poleward at the solar surface and equatorward at the bottom of the convection zone,
plays a crucial role in the transport of magnetic fluxes. We finally point out that various
stochastic fluctuations associated with the dynamo process may play a key role in producing
the irregularities of the sunspot cycle. }


\maketitle

\section{Introduction}\label{sec1}
The 11-year sunspot cycle is one of the most intriguing natural cycles known to mankind. 
Let us begin by looking at Figure~1, which shows
how the number of sunspots seen over the Sun's surface varied with time during the last four
centuries. Galileo and some of his contemporaries were among the first to study sunspots systematically
in the beginning of the seventeenth century by using the newly-discovered telescope. The first
entries in Figure~1 in the years after 1610 are based on the records left by them.
Then there was a period of about 85 years---known as the {\em Maunder minimum}---during which
sunspots were rarely seen.  After that, the sunspot number has gone up and down in a roughly
periodic manner, with a period of about 11 years, although there have been lots of irregularities.

\begin{figure}
    \centering
\includegraphics[width= 0.9\textwidth]{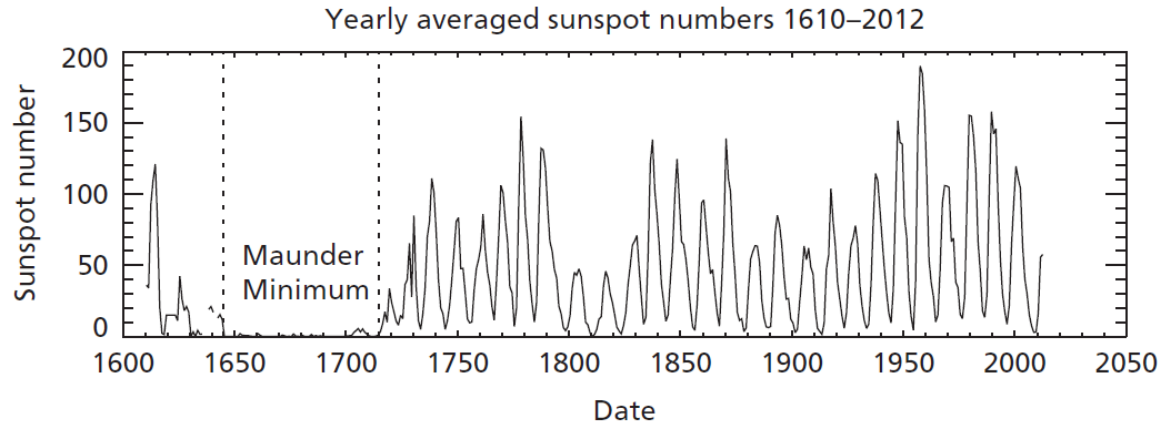}
    \caption{The yearly averaged sunspot number plotted against time since the invention of the telescope.  Credit: David Hathaway.}
\end{figure}

The sunspot cycle was discovered by \citet{Schwabe1844}.  A first clue about the physical nature of
sunspots came with the discovery by \citet{Hale08} of Zeeman splitting in the spectra of
sunspots, from which it can concluded that sunspots are regions of strong magnetic field of about 3000
gauss (or 0.3 tesla)---approximately 5000 times stronger that magnetic field near the 
geomagnetic poles. The discovery of magnetic fields in sunspots
was a momentous discovery in the history of physics, since this was the first time that somebody
conclusively established the existence of magnetic fields outside the Earth's environment.  Now
we know that magnetic fields are ubiquitous in the astronomical universe, with many planets,
stars and galaxies having magnetic fields. 

With the discovery of the magnetic fields in sunspots, it became clear that the 11-year sunspot 
cycle is essentially the magnetic cycle of the Sun.  Since the Sun is made of matter in the
plasma state, one can presume that this magnetic cycle is due to some plasma processes.  The
earliest model of the sunspot cycle given by \citet{Parker55a} and developed further by \citet{SKR66}
is now referred to as the $\alpha \Omega$ dynamo model.  With new observational and theoretical
discoveries coming along, there was need to suitably modify and upgrade this model 
to a more comprehensive
model known as the {\em flux transport dynamo model}. Although there may still be some critics
unwilling to accept the flux transport dynamo as the appropriate theoretical model for the 
sunspot cycle, this model has emerged 
as the currently favoured theoretical model of the sunspot cycle and undoubtedly deserves a
careful consideration.  The present author was lucky that the most crucial years of
growth of this model roughly coincided with his scientific career and our group could make some
key contributions to the development of this model.  

The aim of this presentation
is to introduce the flux transport
dynamo model to plasma physicists who may not have much familiarity with the phenomenology
of the sunspot cycle, highlighting some of the contributions from our group.  This is not
a comprehensive review of the whole field.  The choice of topics has been, to some extent,
guided by the research interests of our group.
We refer the readers to several reviews of the solar dynamo in which the flux transport dynamo
model has been discussed extensively \citep{Charbonneau10,Charbonneau14, Chou11, Karakreview14}. We
may also mention that increasingly more data are coming about spots and cycles of other solar-like
stars, making it clear that the Sun is not an unusual star in having magnetic activities.  Whether
the solar dynamo models can be readily extrapolated to explain the cycles of all solar-like stars is
an important question on which the last word has not been said yet \citep{Chou17}. For a
non-technical introduction to the sunspot cycle and dynamo theory, readers may look at the popular
science book by \citet{Chou15}.

After summarizing the relevant observational data in section~2, we shall discuss the theory of sunpspot
formation in section~3.  Then section~4 will be devoted to introducing the basics of the flux transport dynamo
model and discussing how this model 
explains the regular periodic features of the sunspot cycle. The question of how the irregularities of the sunspot cycles arise
will be briefly discussed in section~5.  Then our conclusions will be summarized in section~6.

\section{Relevant observational data}\label{sec2}

About a decade after Hale's famous discovery of magnetic fields in sunspots \citep{Hale08}, \citet{Hale19}
made another important discovery.  Often two sunspots are seen side by side,
although sometimes both the sunspots in a pair may not be equally well formed. \citet{Hale19}
found that usually the two sunspots occurring in a pair have opposite magnetic polarities. 
The occurrence of such bipolar sunspot pairs suggests the existence of a sub-surface
strand of magnetic flux which presumably
occasionally breaks through the solar surface as shown in Figure~2. If the two intersections
of the strand of magnetic flux with the surface become the two sunspots, then magnetic
field lines would come out of one sunspot (making its polarity positive) and would go down
into the other sunspot (making its polarity negative).

\begin{figure}
    \centering
\includegraphics[width= 0.4\textwidth]{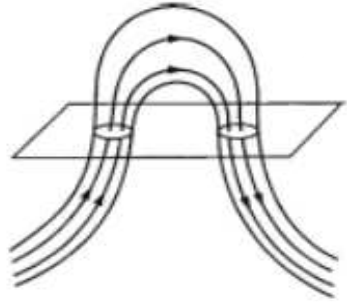}
    \caption{A magnetic flux tube piercing through the solar surface and giving rise to two sunspots of opposite magnetic polarity.}
\end{figure}

Figure~3 shows a magnetogram
map of the Sun in which white and black colours respectively indicate regions of positive and
negative magnetic polarities, whereas grey colour is put in the regions where the magnetic 
field is too weak to be detected by the magnetogram.  In the magnetogram map, a
bipolar sunspot pair appears as a white patch and a black patch side by side. We see
in Figure~3 that the right sunspots in the sunspot pairs in the northern hemisphere are
positive (white patches), whereas the right sunspots in the sunspot pairs in 
the southern hemisphere are
negative (black patches). This is the case for a particular 11-year cycle. 
In the next cycle, the polarity reverses. The
right sunspots in the northern hemisphere would become negative in the next cycle and the
right sunspots in the southern hemisphere would become positive. 

\begin{figure}
    \centering
\includegraphics[width= 0.6\textwidth]{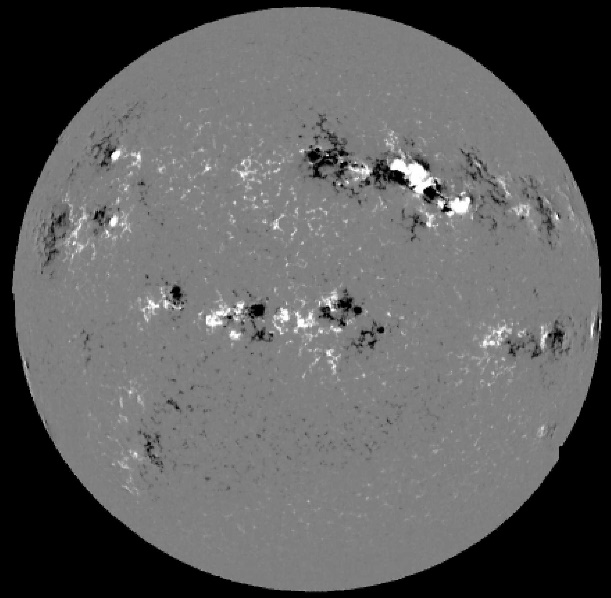}
    \caption{A magnetogram map of the solar disk, with white, black and grey indicating regions of positive, negative and very weak
    magnetic field.}
\end{figure}

We point out another thing in Figure~3. The line joining the centres of the two sunspots in a bipolar sunspot
pair tends to be nearly parallel to the solar equator. Hale's co-worker Joy, however,
noted that there is a systematic tilt of this line with respect to the equator (the right sunspot
in a pair usually appearing closer to the equator) and that this tilt of sunspot pairs increases with latitude 
\citep{Hale19}. This
result is usually known as {\em Joy's law}. The tilts, however, show a considerable amount of
scatter around the mean given by Joy's law. As we shall see later, this law of tilts of sunspot
pairs plays a very important role in solar dynamo theory.

\begin{figure}
    \centering
\includegraphics[width= 0.5\textwidth]{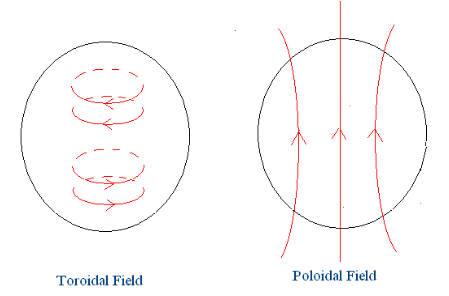}
    \caption{The two possible components of the solar magnetic
    field: (a) the toroidal component and (b) the poloidal component.}
\end{figure}

For the time being, if we ignore the tilts of bipolar sunspots and assume that these appear at the
same latitude, then Figure~3 suggests the possible existence of a sub-subsurface magnetic field
as shown in Figure~4(a).  The magnetic field with this type of configuration is known as the 
{\em toroidal field}. On the other hand, Figure~4(b) shows what is called the {\em poloidal field}---looking
like what the geomagnetic field is expected to be.  Dynamo theory developed historically by
following the mean field approach in which we do an ensemble average of various physical quantities.
When we carry on this kind of averaging, the mean magnetic field can often be regarded as
axi-symmetric, i.e.\ independent of $\phi$ in spherical coordinates.  We can write this mean
magnetic field as 
$$\Bb = B_{\phi} (r, \theta) \ep + \nabla \times [A (r, \theta) \ep]. \eqno(1)$$
The toroidal field is given by $B_{\phi} (r, \theta) \ep$, whereas $\nabla \times [A (r, \theta) \ep]$
gives the poloidal field of which the components are given by
$$B_r =  \frac{1}{r \sin \theta} \frac{\pa}{\pa \theta} (\sin \theta A), \; \;
B_{\theta} =  - \frac{1}{r} \frac{\pa}{\pa r} (r A). \eqno(2)$$
It is easy to check that the contours of constant $r \sin \theta A$ give the magnetic field lines of the
poloidal field in the poloidal plane.

In the fundamental paper on dynamo theory,
\citet{Parker55a} suggested that the sunspot cycle is produced by an oscillation between the toroidal
magnetic field of the Sun
and a poloidal field of the type shown in Figure~4(b).  The proper
observational proof for the existence of such oscillations came
only several decades later, when solar astronomers had gathered sufficient data for the polar
magnetic field of the Sun.  The upper part of Figure~5 shows the temporal variation of the poloidal
field at the two poles of the Sun, whereas the lower part shows the sunspot number, which is a
proxy of the toroidal magnetic field.  Note that the polar field, first discovered by \citet{Babcock55}, is of order a few gauss (keep in mind that 1 gauss = $10^2$ $\mu$T)---much weaker than the 3000-gauss field in the 
interiors of sunspots. We see a clear oscillation between the toroidal and the
poloidal fields.

\begin{figure}
    \centering
\includegraphics[width= 0.8\textwidth]{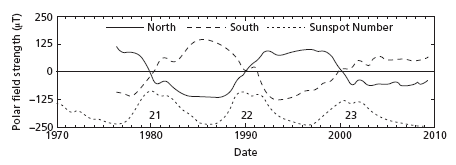}
    \caption{The variation with time of the magnetic field at the two
    poles of the Sun, along with the sunspot number shown at the bottom. Credit: David Hathaway.}
\end{figure}

Let us now say a few words about the appearance of the poloidal field on the solar surface.  At a
particular time, it is found that there would be a latitude belt over which this field at the
surface would have a particular sign.  These latitude belts shift towards the poles with the 
progress of the sunspot cycle.  This is in contrast to sunspots, which appear at lower and lower
latitudes with the progress of the cycle.  Figure~6 is a time-latitude plot in which the shaded
regions indicate the latitudes where sunspots appeared at a particular time, whereas the colours
indicate the longitude-averaged poloidal field at the surface. 
During a sunspot cycle, the shades appear closer to the equator with time, indicating
that sunspots appear at lower latitudes with the progress of the cycle.  On the other hand,
the colours indicating the longitude-averaged poloidal field show a trend of poleward
migration. Since the shaded regions in Figure~6 look like a pattern of repeated
butterflies, they make up what is called the butterfly diagram.
We note that the polar field
is strongest at the time of the sunspot minimum and reverses at the time of the sunspot maximum.
A reversal of the polar field at the time of a sunspot maximum was first observed by \citet{Bab59}.
The theoretical explanation of Figure~6 should be a major goal of a theoretical model of the 
sunspot cycle.

\begin{figure}
    \centering
\includegraphics[width= 0.7\textwidth]{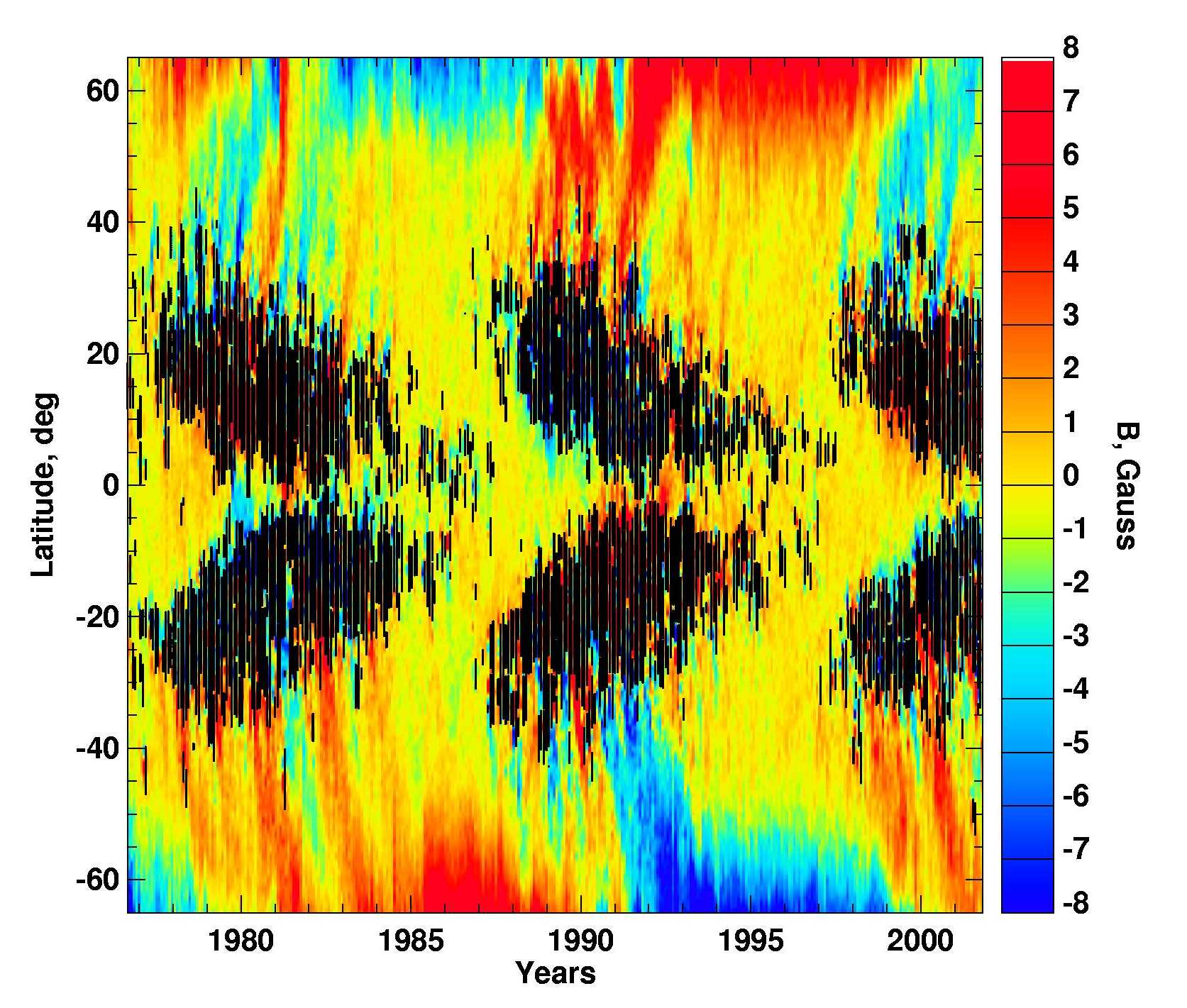}
    \caption{The time-latitude plot of the longitude-averaged radial
    magnetic field at the solar surface superposed on the butterfly
    diagrams of sunspots indicating the latitudes where sunspots are
    seen at a particular time.}
\end{figure}

For the sake of completeness, it may be mentioned that the poloidal field has been found to be
confined in small flux tubes having diameter of the order of a few hundred km with magnetic field
of order 1000 gauss \citep{Stenflo73, Tsuneta08}. Early magnetograms did not resolve these small flux
tubes and gave what would be the average value of the poloidal magnetic field if magnetic fields
inside the flux tubes were spread over the solar surface. The colours in Figure~6 correspond to
such average values.

\section{Theory of sunspot formation}\label{sec3}

Before discussing the theory of how sunspots form, let us make a few comments about the 
interior structure of the Sun according to what is often called the standard model of 
the Sun. The energy generated by nuclear fusion in the central region of the Sun is transported 
outward by radiative transfer till a radius of $0.7 R_{\odot}$, where 
$R_{\odot}$ is the solar radius. The region from $0.7 R_{\odot}$ to the solar surface turns 
out to be unstable to convection, where heat is transported by convection.  It is this 
region, known as the {\em solar convection zone}, where the dynamo action takes place. Sunspots 
are concentrations of magnetic field sitting at the top of this turbulent convection 
zone.

\begin{figure}
    \centering
\includegraphics[width= 0.7\textwidth]{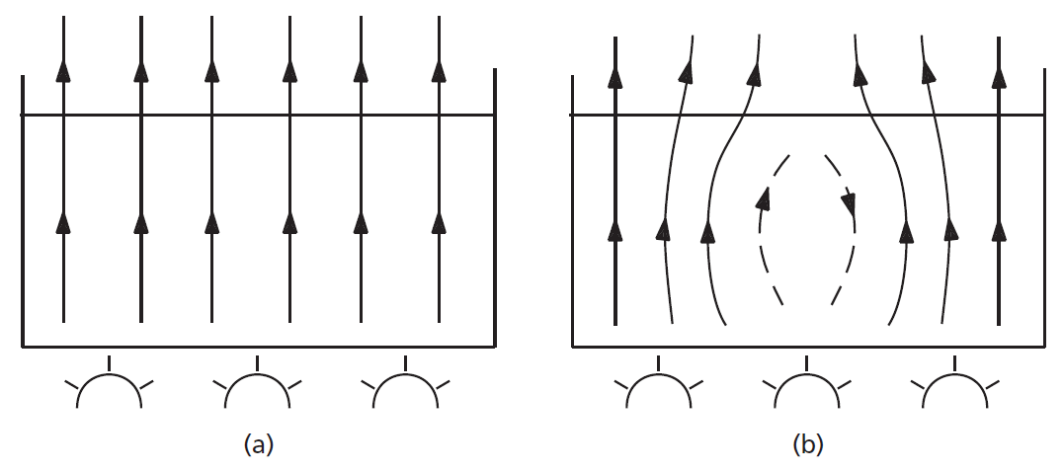}
    \caption{A plasma with vertical magnetic field heated from
    below. (a) The initial configuration. (b) The likely appearance of the magnetic field after convection starts (the dashed lines indicating the convective motions in
    the plasma).}
\end{figure}

Why do magnetic fields remain concentrated within the sunspots instead of filling up 
all space? To address this question, we need to consider the interaction between the 
magnetic field and the convection. This subject is known as {\em magnetoconvection}. 
\citet{Chandra52} worked out the linear theory 
of this subject.  Simulations to study the nonlinear evolution 
of the system were carried out later \citep{Weiss81}.  Consider that a plasma with a vertical 
magnetic field is heated from below, as sketched in Figure 7(a).  The magnetic tension 
tries to oppose convection.  Eventually, when convection is initiated by making the 
vertical temperature gradient sufficiently strong, space gets divided into two kinds 
of regions, as shown in Figure 7(b).  From some regions, the magnetic field lines are expelled 
and convection can take place freely.  In other regions, the magnetic field lines are concentrated 
and convection gets suppressed. Sunspots are presumably such regions of concentrated 
magnetic field---known as magnetic flux tubes---within which the convective heat transport is 
inhibited.  The tops of such regions appear darker compared to the surroundings
because of the decreased heat transport.

\begin{figure}
    \centering
\includegraphics[width= 0.7\textwidth]{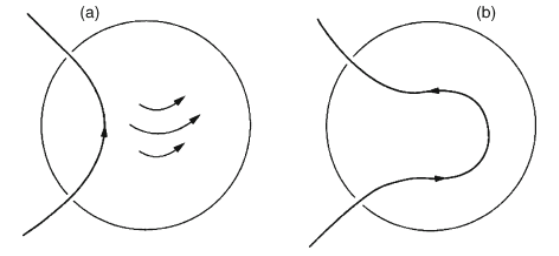}
    \caption{The generation of the toroidal field by the stretching of
    a poloidal field line by differential rotation. (a) An initial poloidal
    field line, with small arrows indicating rotation varying with latitude.
    (b) A sketch of the field line after it has been stretched by the faster
    rotation near the equatorial region.}
\end{figure}

The Sun does not rotate like a solid body, the equatorial region having a higher angular 
velocity.  It has been realized from the early years of MHD research that magnetic field lines would be nearly
``frozen'' in the plasma in a large astrophysical body
like the Sun (see, for example, \citet{Chou98}, section 14.2) and that the differential 
rotation would stretch out a poloidal field to produce a produce a toroidal field, 
as sketched in Figure~8. It should be apparent from this figure that the toroidal field
will have opposite signs in the two hemispheres.  
If parts of this toroidal field rise to the surface, then we would have bipolar sunspot 
pairs with opposite polarity in the two hemispheres, in agreement with observations presented in Figure~3.  \citet{Parker55b} 
realized that the pressure of the magnetic field may cause a region of plasma with strong
magnetic field to expand, giving rise to what is called {\em magnetic buoyancy}. If the
magnetic field exists in the form of a magnetic flux tube, then the pressure balance
condition between its inside and outside would give
$$p_e = p_i + \frac{B^2}{2 \mu_0}, \eqno(3)$$
where $p_e$ and $p_i$ are the gas pressures in the exterior and the interior of the
flux tube, while $B$ is the magnetic field inside the flux tube.  It follows from (3)
that $p_i < p_e$, which implies that the interior and exterior densities also may often,
though not always, satisfy the relation $\rho_i < \rho_e$. We shall not get into a discussion
of the circumstances under which this may happen. If the interior density in some regions of
the flux tube becomes less than the exterior density, then those regions of the flux tube
become buoyant and rise against gravity.
 
\begin{figure}
    \centering
\includegraphics[width= 0.7\textwidth]{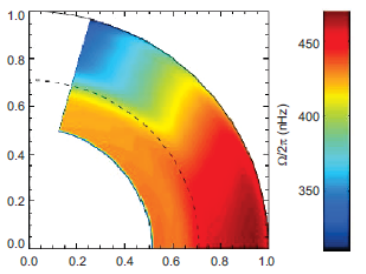}
    \caption{The map of angular velocity distribution in a poloidal
    plane inside the Sun, as obtained by helioseismology. From \citet{Howe05}, as
    given in \citet{Basu16}.}
\end{figure}
 
One of the major breakthroughs in modern astrophysics is that {\em helioseismology}---the study 
of solar oscillations---has succeeded in mapping the differential rotation in the interior 
of the Sun. Figure~9 shows the map of angular velocity distribution
found by helioseismology. It can be seen in this map that the bottom
of the convection zone at $0.7 R_{\odot}$ indicated by the dashed circle is a region of concentrated 
differential rotation, where the angular velocity changes rather
sharply in the radial direction.
We believe that the toroidal magnetic field is primarily generated 
at the bottom of the convection zone and then parts of it rise to the surface due to 
magnetic buoyancy, as shown in Figure~10.  The Coriolis force due to the rotation of 
the Sun may act on the rising part of the flux tube and make it tilted, in accordance 
with Joy’s law.  

\begin{figure}
    \centering
\includegraphics[width= 0.5\textwidth]{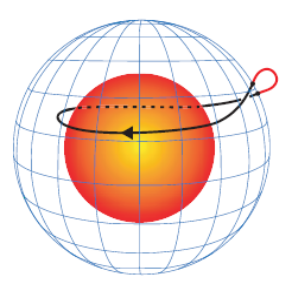}
    \caption{The production of a tilted bipolar sunspot pair on the solar
    surface by the buoyant rise of a part of the toroidal field. From \citet{DG06}.}
\end{figure}

The rise of flux tubes due to magnetic buoyancy was first studied by doing simulations
on the basis of the thin flux tube equation \citep{Spruit81, Chou90}. The first
such simulation in a 2D planar geometry was carried out by \citet{Moreno86}. 
We were the first to study the buoyant rise of flux 
tubes in a spherical geometry by incorporating the Coriolis force 
\citep{Chou89} and gave a theoretical explanation of Joy’s law \citep{Dsilva93} about 
three-quarters of a century after its observational discovery \citep{Hale19}.
We found that the effect of the Coriolis force much stronger than what was suspected before
\citep{CG87} and its effects would have been much larger than what is seen
in observations unless the magnetic field inside the buoyant flux tubes was
sufficiently strong.
By requiring that theory matches observations, we were able to conclude that the magnetic 
field at the bottom of the convection zone has to be as strong as $10^5$ gauss \citep{Dsilva93}. This result was confirmed by the simulations of other groups \citep{Fan93, Caligari95}. As we shall point out in section~4, this value of the toroidal
field was crucial in constraining some aspects of the dynamo process.

Lastly, it may be mentioned that some of the later simulations of the buoyant rise of flux tubes
went beyond the thin flux tube approach and were based on the full MHD equations. Rather than
getting into a discussion of this subject, we refer the interested reader to the excellent
review by \citet{Fan09}. 

\section{The basics of the flux transport dynamo model}

We have discussed how the poloidal field can be stretched by differential rotation 
to produce the toroidal field, from which sunspots form.  To explain the observed 
oscillation between the poloidal and the toroidal fields, we need a mechanism for 
producing back the poloidal field from the toroidal field.  We invoke the idea of 
\citet{Bab61} and \citet{Leighton69} on how the poloidal field can be generated from 
the decay of a tilted bipolar sunspot pair,
like the pair shown in Figure~10. Suppose the sunspot at the 
higher latitude in a pair has positive (negative) polarity.  Typical sunspots live for a few 
days.  When this sunspot pair decays, more positive (negative) polarity is spread 
around at the higher latitude and the opposite polarity at the lower latitude.  What 
we thus get is a poloidal field.  This {\em Babcock--Leighton mechanism} for the generation 
of the poloidal field is somewhat different from the $\alpha$-effect \citep{Parker55a, SKR66}, in which helical turbulence twists the toroidal 
field to produce the poloidal field. If the toroidal field is as strong $10^5$ gauss, 
as suggested by buoyancy simulations, then such twisting is not possible and we 
believe that the Babcock--Leighton mechanism is the dominant mechanism for 
generating the poloidal field in the Sun. It is possible that the $\alpha$-effect
is also present in the regions of weak magnetic field along with the dominant
Babcock--Leighton mechanism.  In fact, the $\alpha$-effect may be necessary for the
dynamo to recover from grand minima like the Maunder minimum when sunspots disappear
and the Babcock--Leighton mechanism may be absent \citep{KarakChou13,Passos14}.  

We now can think of a minimalistic dynamo model in which the toroidal and poloidal 
magnetic fields sustain each other through a feedback loop: differential rotation 
producing the toroidal field from the poloidal field and the Babcock-Leighton 
mechanism producing the poloidal field back from the toroidal field (which is responsible 
for the tilted bipolar sunspots). It turns out that this minimalistic model leads 
to a dynamo wave propagating poleward, in accordance with what is known as the
Parker--Yoshimura sign rule \citep{Parker55a, Yoshimura75}.  This means 
that sunspots should appear at higher latitudes with the progress of 
the cycle, opposite of what is observed \citep{CSD95}.  
It is clear that we need something else to turn things around.
The meridional circulation of the Sun to be discussed in the next
paragraph turns out to be this `something else'.

The Sun is known to have a plasma flow at the surface from the equator to the 
poles, the maximum amplitude of the flow at mid-latitudes being about 20 m s$^{-1}$.  
As we do not expect the plasma to pile up near the solar poles, this poleward
plasma flow must be a part of a larger meridional circulation pattern having an
equatorward return flow somewhere below the surface bringing back the plasma
that has flown to the poles.  
Since the turbulent stresses in the convection zone are responsible the large-scale flows 
there \citep{Chou21a}, we expect the meridional circulation to be confined within the convection 
zone.  Dynamo models are found to give best results if the return flow of the meridional 
circulation is at the bottom of the convection zone, although some studies have 
been done with more general kinds of flows \citep{HKC14}. 
Within the last few years, helioseismology has confirmed the existence of 
the return flow at the bottom of the convection zone \citep{Raja15, Giz20}, validating different theoretical groups who have been 
constructing dynamo models assuming such a flow for many years.

\begin{figure}
    \centering
\includegraphics[width= 0.6\textwidth]{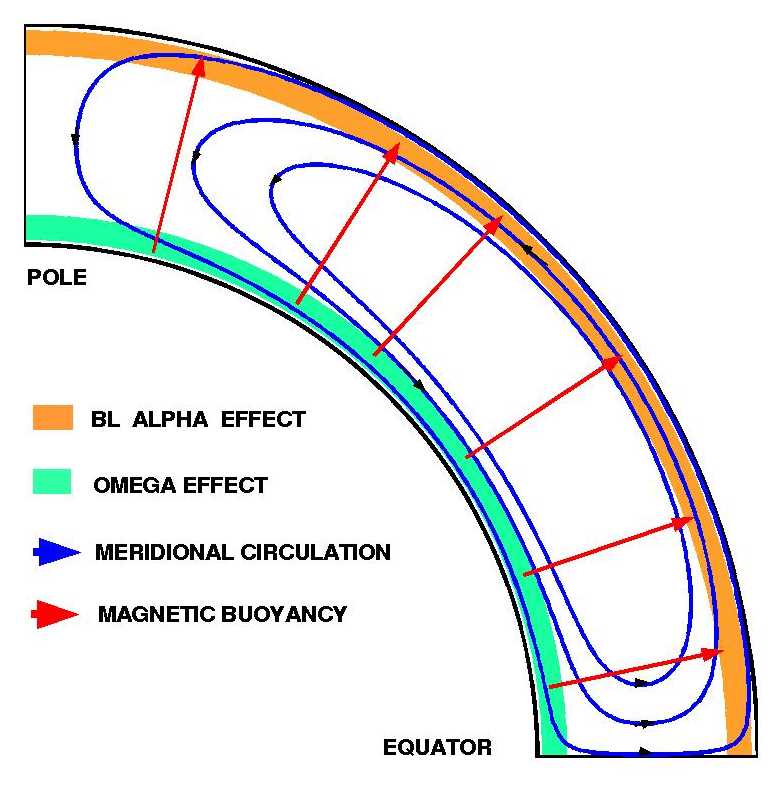}
    \caption{A cartoon indicating the essential ingredients of
    the flux transport dynamo model.}
    \label{FTD}
\end{figure}

Figure 11 shows a cartoon summarizing what is called the {\em flux transport dynamo 
model}. The green colour indicates the region at the bottom of the convection 
zone where differential rotation produces the strong toroidal field.  The red 
arrows represent magnetic buoyancy due to which the toroidal field rises to 
the solar surface to produce sunspots.  The brown region near the surface is 
where the poloidal field is produced by the Babcock-Leighton mechanism.  The 
blue curves indicate the all-important meridional circulation. The toroidal 
field produced at the bottom of the convection zone is advected by the meridional 
circulation there, ensuring that sunspots appear closer to the equator with 
the progress of the sunspot cycle. On the other hand, the poleward meridional 
circulation near the surface advects the poloidal field generated there, in 
agreement with observations. 

Although some basic ideas of this model
were suggested on the basis of a 1D model in an early paper by
\citet{Wang91}, the first 2D calculations based on this model 
were done in the mid-1990s \citep{CSD95, Durney95}.
The paper by \citet{CSD95}, perhaps the most important
paper in the present author's scientific career, convincingly showed that the
meridional circulation can indeed turn things around and make sunspots appear
at lower latitudes with the progress of the cycle rather than at higher latitudes
which would happen in the absence of the meridional circulation.  This established
the flux transport dynamo as a promising model of the solar dynamo and several
groups started studying different aspects of this model within the next few years \citep{Durney97, DC99, Kuker01, DG01, Nandy01, Nandy02, Bon02, Guerrero04, CNC04}.

The flux transport dynamo model was first developed by doing calculations with 
the help of mean field equations obtained by averaging over the turbulence 
in the convection zone. The mean field equation for the evolution of the
magnetic field is the famous {\em dynamo equation}
$$\frac{\pa \Bb}{\pa t} = \nabla \times (\vb \times \Bb) + \nabla \times (\alpha \Bb) 
+ \lambda_T \nabla^2 \Bb, \eqno(4)$$
where $\lambda_T$ is the turbulent
diffusivity inside the convection zone and $\alpha$ is the parameter which
governs the generation of the poloidal field (see, for example, \citet{Chou98}, Chapter~16).
The parameter $\alpha$ for the Babcock--Leighton mechanism has to be specified somewhat
differently compared to the classical $\alpha$-effect
proposed by \citet{Parker55a} and \citet{SKR66}.  The mean magnetic field written
in the form (1) can be substituted in (4), whereas the velocity has to be written as
$$\vb = \vf + r \sin \theta \, \Omega (r, \theta) {\bf e}_{\phi},\eqno(5)$$ 
where $\Omega (r, \theta)$ is the angular velocity in the interior of the
Sun and $\vf$ is the velocity of meridional circulation having components
in $r$ and $\theta$ directions.  On substituting (1) and (5) into (4), some
reasonable assumptions lead to the following coupled equations for the poloidal
and the toroidal fields
$$
\frac{\pa A}{\pa t} + \frac{1}{s}(\vf.\nabla)(s A)
= \lambda_T \left( \nabla^2 - \frac{1}{s^2} \right) A + \alpha B,
\eqno(6)$$
$$ \frac{\pa B}{\pa t} 
+ \frac{1}{r} \left[ \frac{\pa}{\pa r}
(r v_r B) + \frac{\pa}{\pa \theta}(v_{\theta} B) \right]
= \lambda_T \left( \nabla^2 - \frac{1}{s^2} \right) B 
+ s(\Bf_p.\nabla)\Omega + \frac{1}{r}\frac{d\lambda_T}{dr}
\frac{\partial}{\partial{r}}(r B), \eqno(7)$$
where $s = r \sin \theta$ and $\Bf_p$ is the poloidal field with
the components given by (2). To understand the behaviour of the flux
transport dynamo, we have to solve (6) and (7) simultaneously after
suitably specifying the various parameters $\Omega$, $\vf$, $\lambda_T$
and $\alpha$.
  
\begin{figure}
    \centering
\includegraphics[width= 0.7\textwidth]{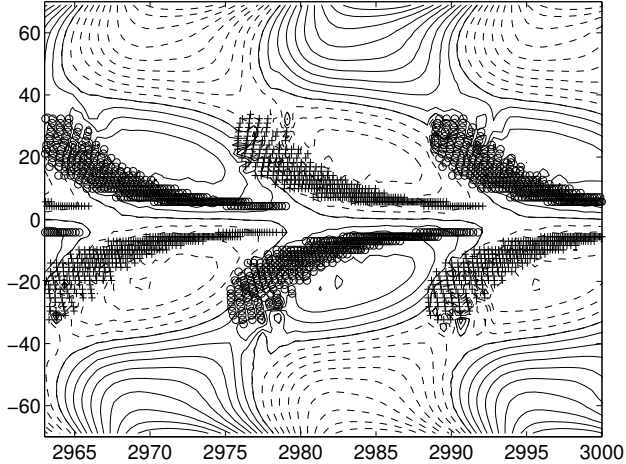}
    \caption{A theoretical time-latitude plot from \citet{CNC04} based on their dynamo calculation. The shaded regions indicate the latitudes where sunspots are seen at different times, whereas the contours indicate the values of the radial magnetic field at the solar surface.}
    \label{FTD}
\end{figure}  
  
To get an idea about the nature of the solutions of these  equations,
look at Figure~12 taken from \citet{CNC04}. It 
is a theoretical time-latitude plot to be compared with the observational 
plot shown in Figure~6.  The shaded regions in both the plots indicate sunspots. 
The contours of constant radial field at the surface seen Figure~12 have to be 
compared with the colours in Figure~6.  Given the fact that Figure~12 was about 
the first such theoretical plot produced from the flux transport dynamo model, 
hopefully all readers will agree that the fit between theory and observations was 
impressive.

We have pointed out in section~1 about the evidence of magnetic cycles in other solar-like
stars.  The difficulty of making models of these stellar dynamos is that we do not
have any detailed information about the differential rotation $\Omega$ or the
meridional circulation $\vf$ for other stars besides the Sun.  However, these large-scale fluid flows
can be calculated from mean field models of the large-scale flows
\citep{Kitchatinov95, Kitchatinov12} and then used for constructing flux
transport dynamo models of solar-like stars \citep{KKC14}.  
Such models can match many aspects
of observational data.  However, there are still doubts whether the flux transport
dynamo model is universally applicable to all solar-like stars \citep{Chou17}.

\section{Modelling irregularities of the sunspot cycle} 
 
We have given some idea of how the regular periodic features of the sunspot cycle 
are explained with the flux transport dynamo model.  A look at
Figure~1 makes it clear that the sunspot cycle is only
approximately periodic. We shall now turn our attention
to the question of how the flux transport dynamo 
model can be applied to explain the observed
irregularities of the sunspot 
cycle.  See a review by \citet{Chou14} on this subject. 
An early idea was that the irregularities of
the sunspot cycle are manifestations of chaos 
resulting from the nonlinearities of the dynamo problem \citep{Weiss84}.  
However, it appears that the most obvious kinds of nonlinearities would not produce 
the sustained irregularities we observe, and some stochastic fluctuations may be the 
more appropriate cause of the cycle irregularities \citep{Chou92, Hoyng93}.  
We should point out that there are some irregularities which possibly result from
nonlinear chaos.  For example, the Gnevyshev--Ohl effect---the fact that the
odd-numbered cycle was stronger than the previous even-numbered cycle for several cycles---is
probably a manifestation of nonlinear chaos \citep{CSZ05, Char07}. Since stochastic fluctuations
are likely to be the more important source for sunspot cycle irregularities, let
us now discuss how these fluctuations arise.

As we have pointed out, a crucial ingredient of the flux transport dynamo is
the Babcock--Leighton mechanism, which involves the tilts of bipolar sunspot pairs
as given by Joy's law.  However, Joy's law happens to be a law of statistical
averages, there being a random distribution of tilt angles around the average given by
Joy's law \citep{SK12}. This randomness presumably arises due to the
effect of turbulence on the magnetic flux tubes rising through the convection
zone \citep{Longcope02} and introduces
fluctuations in the Babcock--Leighton mechanism. \citet{CCJ07} identified
these fluctuations in the Babcock--Leighton mechanism arising out
of the randomness in the sunspot tilt angles as the main source of irregularities
in the sunspot cycle and developed a method for predicting the next upcoming cycle.

Different methods have been suggested over the years for the prediction of a sunspot
cycle before its onset.  After the development of the flux transport dynamo model, 
whether it is possible to predict a future cycle based on this theoretical dynamo
model became an important question.  \citet{DG06} proposed a method of making such
predictions which suggested that the upcoming cycle 24 would be a very strong cycle.
\citet{CCJ07} and \citet{Jiang07} pointed out several logical flaws in the Dikpati--Gilman arguments
and developed an alternative methodology for predicting the next cycle based on the
version of the flux transport dynamo model they had developed.
They predicted a rather low value for the
peak of the sunspot cycle 24, which turned out to be the first successful dynamo-based
prediction of a sunspot cycle before its onset. Figure~13 is a plot of observed sunspot
numbers along with the two theoretical predictions of cycle 24 due to \citet{DG06} and \citet{CCJ07}.  

\begin{figure}
    \centering
\includegraphics[width= 0.8\textwidth]{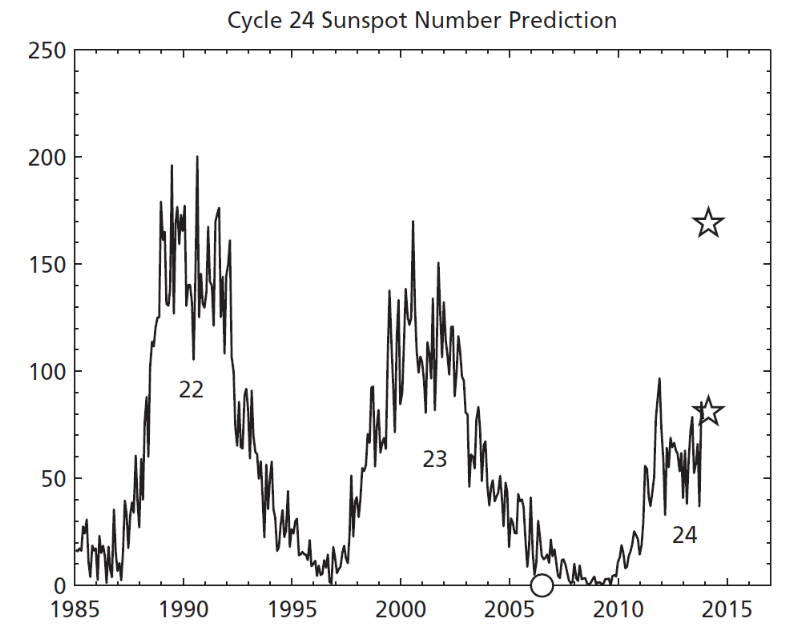}
    \caption{The predictions for the peak of cycle 24 by
    \citet{DG06} and \citet{CCJ07} are indicated respectively
    by the upper star and the lower star in this plot showing
    the variation of sunspot number with time during that era. The circle of the horizontal axis indicates the time when
    these predictions were made.}
    \label{FTD}
\end{figure}  

We now realize that, apart from the fluctuations in the 
Babcock--Leighton mechanism, fluctuations in the meridional
circulation also can be an additional source of irregularities
in the sunspot cycles.
The period of the flux transport dynamo depends on the strength of the meridional
circulation, periods becoming shorter when the circulation is stronger \citep{DC99, Nandy01}. If there are fluctuations in the meridional
circulation, that would certainly introduce irregularities in the sunspot cycle \citep{Karak10}. Especially, durations of different cycles would vary due to such fluctuations.
From the historical data of past cycles, \citet{KarakChou11} indeed found indirect
evidence for fluctuations in the meridional circulation in the past.  Diffusion acting
on longer cycles may make them weaker, giving rise to an anti-correlation between the
cycle duration and strength. This anti-correlation readily leads to an explanation of 
the Waldmeier effect that stronger cycles rise faster, because they 
are shorter due to this anti-correlation \citep{KarakChou11}. 

Taking the fluctuations in the Babcock--Leighton mechanism and the fluctuations in the
meridional circulation as the two main sources of irregularities in the sunspot cycles,
\citet{CK12} developed a comprehensive model of grand minima like the Maunder minimum
during the seventeenth century, which can be seen in
Figure~1.  There is now indirect evidence (from the analysis
of polar ice cores) that there have been about 27 grand minima in the last 11,000 yr
\citep{Uso07}. The results of \citet{CK12} are in broad agreement with this. With
the realization that the fluctuations in the meridional circulation are so important
in producing irregularities in the sunspot cycle, it is clear that these fluctuations
have to be taken into consideration along with the fluctuations in the Babcock--Leighton
mechanism for the prediction of future cycles.  How this can be done has been
discussed by \citet{HC19}.

Lastly, we may mention that two papers from our group---\citet{CCJ07} 
on the prediction of sunspot
cycles and \citet{CK12} on the grand minima---were selected as 
“Editors’ suggestion” in {\em Physical Review Letters}, showing that the subject of
irregularities of the sunspot cycle is of considerable interest to the physics community.  

\section{Concluding remarks}

The basic idea of the solar dynamo is that the sunspot cycle is produced by an
oscillation between the toroidal and poloidal components of the solar magnetic
field.  The toroidal field, from which the sunspots arise due to magnetic buoyancy,
is generated by the stretching of the poloidal field by differential rotation. How
the poloidal field arises from the toroidal field is less certain.  The current
flux transport dynamo models invoke the Babcock--Leighton mechanism in the place
of the older $\alpha$-effect which can work only if the toroidal field is much
weaker what we now believe it to be.  The meridional circulation, which is 
poleward at the surface and is expected to be equatorward at the bottom of the
convection zone, plays a crucial role in the flux transport dynamo model in ensuring
that the solar magnetic fields are transported in agreement with observations. We 
have also discussed how the irregularities in the Babcock--Leighton mechanism and
in the meridional circulation may give rise to the observed irregularities in the
sunspot cycle.

The flux transport dynamo model developed historically by following the mean
field approach, in which we average over turbulence in the convection zone. Equation
(4), which leads to Equations (6) and (7), is based on such mean field approach.
Using the observational input for the solar differential rotation $\Omega$ and the
meridional circulation $\vb_m$ in Equation (5), we can specify the velocity field
$\vb$ and follow the kinematic approach of solving our equations only for the 
magnetic field. As we have already pointed out in section~4, we have to go beyond
the kinematic approach in modelling stellar dynamos for which we do not have
information about the large-scale flows.  Even for the Sun, we have to go beyond the
kinematic approach if we want to study the observed variations of the large-scale flows
with the cycle due to the back-reaction of the dynamo-generated magnetic field 
\citep{Rempel06, Chou21a}. How the observed cyclic variation of the differential rotation,
known as torsional oscillations, arises in the flux transport dynamo model has
been studied by \citet{CCC09}.  The meridional circulation is found to
become weaker at the time of the sunspot maximum \citep{Hathaway10b}.
While the variation of meridional circulation can be included in the dynamo model by
introducing a simple quenching by the magnetic field \citep{KarakChou12}, a proper
theory requires the solving of the equation for \MC\ along with the dynamo
equations \citep{HC17}. It may noted that the mean field models of
the \DF\ and the \MC\ show them to be intimately connected with each other.  For
example, to explain the shear layer of \DF\ just below the solar surface seen in Figure~9, we need
to analyze the thermal wind balance equation arising in the theory of the \MC\
\citep{Chou21b, Jha21}.

Within the last few decades, tremendous advances have been made in the direct
numerical simulation (DNS) of the geodynamo starting with the path-breaking work
by \citet{Glatz95}. While the mean field models played a historically
important role in the growth of the flux transport dynamo model, often a question is
asked whether these models still remain relevant in the present era of DNS.  Indeed,
some impressive DNS calculations of the solar dynamo have been done from around 2010 onwards \citep{Ghi10,Brown10}.  However,
simulating the solar dynamo is much more challenging than simulating the geodynamo
because of the wide ranges of length and time scales involved.  Also, the stratification
of the convection zone within which the density and pressure vary by several orders
of magnitude makes realistic simulations very difficult.  Additionally, in a kinematic
mean field model, one can specify the \DF\ and \MC\ from observational data.  In
contrast, in a DNS, these large-scale flows have to come out of the simulations and,
until one gets the large-scale flows correctly, there is no hope of buidling a
realistic model of the dynamo.  Because of these reasons, simulations of the solar
dynamo are still of rather exploratory nature.  We still have to depend on the mean
field model for providing detailed explanations of different aspects of observational
data. 

One limitation of 2D mean field models is that the Babcock--Leighton mechanism is
an inherently 3D mechanism and can be included in 2D models only through crude
approximations.  In fact, there has been some debate about the best way handling
the Babcock--Leighton mechanism within the framework of 2D models
\citep{Durney97, Nandy01, Munoz10}.  Rather than
going all the way to full 3D simulations, one can think of constructing 3D kinematic
models \citep{YM13,MD14,HCM17}.  In such models, the large-scale flows are specified on the basis of the
observational data, whereas the evolution of the magnetic field is treated in 3D
so that the Babcock--Leighton mechanism is modelled realistically by considering
the decay of tilted bipolar sunspots.

While 2D kinematic mean field models have provided explanations for many aspects
of the sunspot cycle, we certainly need to go beyond them if we want our solar
dynmao models to be sufficiently realistic.  Perhaps 3D kinematic models happen
to be the next important step.  Ultimately our goal should be to carry on 3D simulations
of the large-scale flows and the solar dynamo together.  However, we probably have
to wait for a few years before fully realistic simulations of this kind can be
carried out in a self-consistent manner.

\section{A personal note}

Let me end by mentioning that this paper is based on my 
lecture given on the occasion of the Subrahmanyan Chandrasekhar Prize of
Plasma Physics
being bestowed on me. A Prize named after Professor Chandrasekhar has a special personal
significance for me, since I am presumably the first recipient of this Prize who had Chandrasekhar himself as a professor
at the University of Chicago. As a first-year graduate student there, I took a course on general
relativity taught by him. 

\begin{figure}
    \centering
\includegraphics[width= 1.0\textwidth]{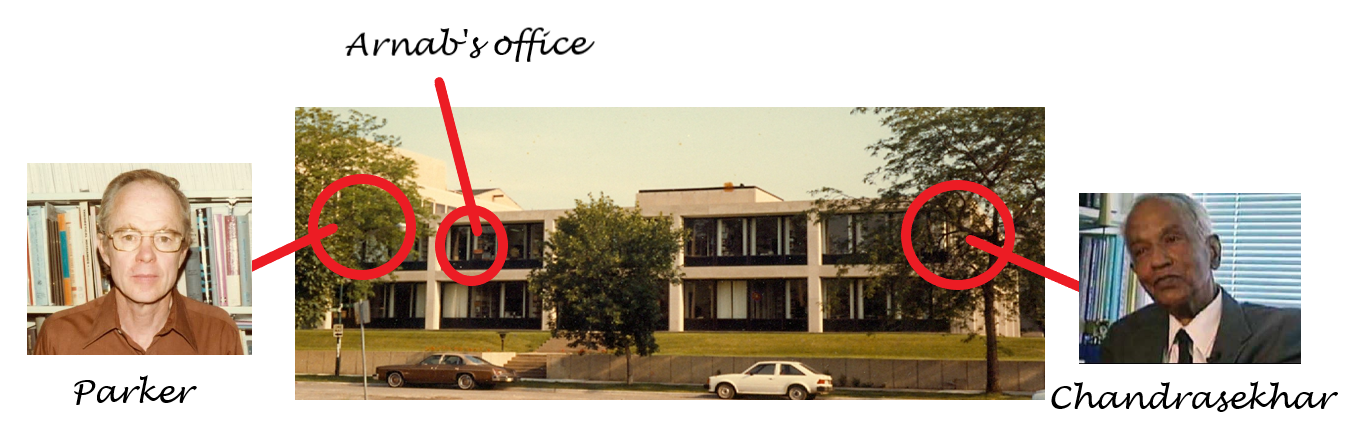}
    \caption{The Laboratory for Astrophysics and Space Research (LASR) in the University of Chicago campus, as
    it was in the 1980s. The upper left corner room was Parker's office, whereas the room next to it was my office.  The upper right corner room was Chandrasekhar's office.}
    \label{FTD}
\end{figure}  

I dedicate this paper to the memory of my PhD supervisor E.N.\ (Gene) Parker, who first initiated me to the field of plasma
astrophysics. He passed away a few months ago. Figure~14 shows the Laboratory for Astrophysics
and Space Research in the University of Chicago campus, where we all had our offices.  The left
corner room in the upper floor was Parker's office and the right corner room was Chandrasekhar's 
office. My office was next to Parker's office.  Working in that office for four years, I had the 
rare privilege of observing these two giants of theoretical astrophysics closely.

Whatever little I have achieved in science during the last few years has been possible because
of the succession of exceptionally brilliant students who decided to work under my supervision
for their PhD---Sydney D'Silva, Mausumi Dikpati, Dibyendu Nandy, Piyali Chatterjee, Jie Jiang,
Bidya Karak, Gopal Hazra.  I am grateful to my colleague Rahul Pandit who insisted on 
nominating me for the Chandrasekhar Prize against my initial hesitation.  I thank Eric
Priest, Kazunari Shibata, Paul Charbonneau, Jie Jiang and Durgesh Tripathi for their
support letters.

Apart from those who helped me professionally in my scientific career, my journey has been
possible due to the support and encouragement of many others---family members, teachers,
friends.  My parents encouraged me from my childhood to take up an academic career. I 
would not have reached here today without the strong support of my wife Mahua.

\bibliography{myref}

\end{document}